\documentclass[a4paper,11pt]{article}
\pdfoutput=1 
\usepackage{jheppub} 
\usepackage[T1]{fontenc} 
\usepackage{multirow}

\newcommand{\f}{\frac}
\newcommand{\lt}{\left}
\newcommand{\n}{\nonumber}
\newcommand{\p}{\partial}
\newcommand{\dd}{{\rm d}}
\newcommand{\rt}{\right}

\newcommand{\arxgr}[1]{\href{http://arxiv.org/abs/#1}{{\ttfamily arXiv:#1[gr-qc]}}}
\newcommand{\arxth}[1]{\href{http://arxiv.org/abs/#1}{{\ttfamily arXiv:#1[hep-th]}}}

\newcommand{\Arxgr}[1]{\href{http://arxiv.org/abs/gr-qc/#1}{{\ttfamily arXiv:#1[gr-qc]}}}
\newcommand{\Arxth}[1]{\href{http://arxiv.org/abs/hep-th/#1}{{\ttfamily arXiv:#1[hep-th]}}}

\title{\boldmath Hawking--Page phase transitions of the black holes in a cavity}
\author[a]{Wen-Bo Zhao,}
\author[a]{Guo-Rong Liu,}
\author[a,1]{and Nan Li \note{Corresponding author.}}
\affiliation[a]{Department of Physics, College of Sciences, Northeastern University \\ No. 3-11, Wenhua Road, Shenyang, 110819, China}
\emailAdd{linan@mail.neu.edu.cn}

\abstract{The Hawking--Page phase transitions of the Schwarzschild and charged black holes are investigated in an extended phase space, in which the black holes are enclosed in a spherical cavity of radius $r_B$ in asymptotically flat space. An effective thermodynamic volume $V=4\pi r_B^3/3$ is introduced for the black hole, and an effective pressure $p$ is defined as the conjugate variable of $V$. The phase transition temperature $T_{\rm HP}$ and the Gibbs free energy $G$ are systematically studied in a grand canonical ensemble with fixed electric potential $\Phi$, and $T_{\rm HP}$ is found to increase with $p$ and decrease with $\Phi$. If the phase transition occurs, $\Phi$ must have an upper bound, such that the cavity radius is always larger than the black hole horizon radius. These phase transition behaviors are further compared to those in the anti-de Sitter space, and the remarkable similarities and notable differences are also discussed in depth. Our work reveals the relationship of the thermodynamic properties of black holes and their specific boundary conditions in different extended phase spaces.}

\begin{document}
\maketitle
\flushbottom

\section{Introduction} \label{sec:intro}

During the last half century, black hole thermodynamics has been one of the most important and fruitful research areas in theoretical physics, and has been continuously offering profound and fundamental insights to our understanding of quantum gravity. The establishment of the four laws of black hole thermodynamics \cite{law} explicitly indicates that a black hole is not simply a mathematical singularity, but should rather be considered as a complicated thermodynamic system with temperature and entropy \cite{Bekenstein}.

However, despite the striking resemblance between black hole thermodynamics and traditional thermodynamics, there remain two obvious discrepancies between them. One is that there is no $p$--$V$ term in the first law of black hole thermodynamics, and the other is that a black hole in asymptotically flat space has negative heat capacity and is thus thermodynamically unstable. Therefore, when a black hole radiates via the Hawking mechanism \cite{rad}, it can no longer maintain thermal equilibrium with the environment and will evaporate eventually. These two aspects make a black hole still somehow different from a usual thermodynamic system (e.g., a $p$--$V$--$T$ system). In fact, both these two aspects can be tackled to be consistent with traditional thermodynamics, and the idea is just to impose appropriate boundary conditions and to introduce an effective pressure $p$ and an effective volume $V$ to the black hole system. In this way, two new dimensions are added into the black hole phase space, and such theories are thus named as black hole thermodynamics in the extended phase space.

Generally speaking, there are two natural choices of the boundary conditions and the ways to restore the $p$--$V$ term. The first is to confine a black hole in the anti-de Sitter (AdS) space, and an effective pressure can be introduced as \cite{Kastor, Dolan}
\begin{align}
p=-\f{\Lambda}{8\pi}, \n
\end{align}
where $\Lambda$ is the cosmological constant. Furthermore, if $\Lambda$ is allowed to vary, $p$ will behave as a thermodynamic variable rather than a fixed background, and an effective volume of a black hole can be defined as the conjugate quantity of $p$ as $V=({\p M}/{\p p})_S$, with $M$ being the black hole mass \cite{KM}. By this means, the first law of thermodynamics is consistent with the Smarr relation, and the analogy between black hole thermodynamics in the extended phase space and traditional thermodynamics becomes more complete. Moreover, it is found that there are two branches of black hole solutions in the extended phase space, a stable large black hole with larger event horizon radius and positive heat capacity and an unstable small black hole with smaller event horizon radius and negative heat capacity, respectively. The former possesses an equation of state very similar to that of a non-ideal fluid and exhibits rich thermodynamic behaviors, such as various critical phenomena. All these relevant topics have been extensively studied in the literature in recent years (see Ref. \cite{rev} for a review and the references therein).

The essential reason that a black hole in the AdS space is thermodynamically stable is based upon the fact that the gravitational potential in the AdS space increases at large distances, acting as a box of finite volume. Therefore, the second way to restore the $p$--$V$ term and to stabilize the black hole is just to enclose the black hole in asymptotically flat space in a cavity, on the wall of which the metric is fixed (i.e., with the Dirichlet boundary condition). Thus, the wall of cavity, as a reflecting boundary, will equivalently play a role of the AdS space. In a recent work \cite{Wang}, the authors considered a new extended phase space for the black hole in a cavity. In a spherical cavity of radius $r_B$, an effective volume of the black hole is first introduced as
\begin{align}
V=\f{4\pi}{3}r^3_B, \n
\end{align}
and $r_B$ is regarded as a thermodynamic variable. Next, an effective pressure is defined as the conjugate quantity of $V$ as $p=-({\p E}/{\p V})_{S}$, with $E$ being the thermal energy of the black hole. It is interesting to see that the order of introduction of the $p$--$V$ term in a cavity is opposite to that in the AdS space, but the thermodynamic properties of black holes are strikingly analogous in these two different extended phase spaces. For example, there are also two black hole solutions, and the large black hole is thermodynamically stable and has an equation of state like a non-ideal fluid \cite{Wang}.

The explorations of the black holes in a cavity had a long history. In Ref. \cite{York}, via the approach of the Euclidean action, York studied the phase structures and transitions of the Schwarzschild black hole in a cavity and in the AdS space, and found remarkable similarities in between. This observation was later confirmed for the charged black hole in a canonical ensemble \cite{Carlip, Lundgren} and in a grand canonical ensemble \cite{Braden}. Later on, similar investigations were extended to other black holes \cite{Parentani:1994wr, Akbar:2004ke, 29, 30, Dias11, Dias12, Wang:2019urm, 35, 36, 37, Tzikas}, the black branes \cite{Emparan:2012be, 21, 22, 23, 24, 25, 26}, the boson stars \cite{31, 32}, and the stabilities of stars, solitons, and black holes in a cavity \cite{Gregory:2001bd, Dolansss, 27, Ponglertsakul1, 28, Ponglertsakul2, Sanchis-Gualhhh}. Simultaneously, some dissimilarities in the phase structures were discovered between the Born--Infeld black hole in a cavity and their AdS counterparts \cite{Wang:2019kxp, Liang:2019dni}. Similarly, the thermodynamic geometry and the weak cosmic censorship conjecture were also found to be different between the cavity and the AdS cases \cite{Wang:2019cax, Wang:2020osg}. Altogether, the similarities in the relevant works further supported the analogy between the cavity and the AdS cases; meanwhile, the discrepancies revealed the potential relations between the thermodynamic properties of black holes and the specific boundary conditions.

Actually, both in the AdS space and in a cavity, the precondition of the existence of black hole solution is that the black hole temperature must be higher than some value. Below this temperature, there will be no black hole solution any longer, and the black hole will turn into the thermal gas. In the AdS space, this process is the famous Hawking--Page (HP) phase transition. It was first investigated for the Schwarzschild black hole in Ref. \cite{HP} and then for the charged black hole in Refs. \cite{Chamblin1, Chamblin2, Peca}. The HP phase transitions have been broadly studied in the literature \cite{Witten, Cald, Birmingham:2002ph, Herzog:2006ra, Cai:2007wz, Nicolini:2011dp, Eune:2013qs, Adams:2014vza, Banados:2016hze, Czinner:2017tjq, Aharony:2019vgs, Mejrhit:2019oyi, Lala:2020lge}, especially in the extended phase space in the AdS space \cite{Spallucci:2013jja, Altamirano:2014tva, Xu:2015rfa, Maity:2015ida, Liu:2016uyd, Hansen:2016ayo, Sahay:2017hlq, Mbarek:2018bau, sanren, HPwo, Astefanesei:2019ehu, Wu:2020tmz, DeBiasio:2020xkv, Wei:2020kra, Li:2020khm, Wang:2020pmb, Belhaj:2020mdr, ydw}. Consequently, it is quite natural to reconsider it in the new extended phase space in a cavity. This issue was briefly mentioned in Ref. \cite{Wang}, with some graphical illustration but no analytical result. Therefore, the purpose of our present work is to provide a thorough exploration of the HP phase transitions of the black holes in a cavity. The HP temperature, the minimum temperature, and the Gibbs free energy of the black holes will be systematically studied. Besides numerical methods, all analytical results will be given. The comparison with the AdS case will also be shown when necessary, so that we can have a global understanding of the HP phase transitions in different extended phase spaces.

This paper is organized as follows. In Sect. \ref{sec:cavity}, we list the thermodynamic properties of the black holes in a cavity and discuss the HP phase transition in detail. In Sect. \ref{sec:HP}, we explore the HP phase transitions of the neutral and charged black holes. The similarities and differences with the results in the AdS space are also presented. We conclude in Sect. \ref{sec:con}. In this paper, we work in the natural system of units and set $c=G_{\rm N}=\hbar=k_{\rm B}=1$.

\section{Black hole thermodynamics in a cavity} \label{sec:cavity}

In this section, we first outline black hole thermodynamics in the extended phase space in a cavity and then show the criterion of the HP phase transition.

\subsection{Extended phase space in a cavity} \label{sec:excavity}

We start from the action of the four-dimensional Einstein--Maxwell gravity,
\begin{align}
I=\f{1}{16\pi}\int\dd^4x\,\sqrt{-g}(R-F_{\mu\nu}F^{\mu\nu}), \n
\end{align}
where $R$ is Ricci scalar and $F_{\mu\nu}$ is the electromagnetic field tensor. Then, the metric of the charged black hole (i.e., the Reissner--Nordstr\"{o}m black hole) reads
\begin{align}
\dd s^2=-f(r)\,\dd t^2+\f{\dd r^2}{f(r)}+r^2\,\dd\theta^2+r^2\sin^2\theta\,\dd\phi^2,\n
\end{align}
where $f(r)=1-{2M}/{r}+{Q^2}/{r^2}$, and $M$ and $Q$ are the mass and electric charge of the black hole, with the electric potential being $A=A_t(r)\,\dd t=-Q\,\dd t/r$. The event horizon radius $r_+$ can be fixed as the largest root of $f(r_+)=0$, so $r_+=M+\sqrt{M^2-Q^2}$. By this means, $f(r)$ can be reexpressed as
\begin{align}
f(r)=\lt(1-\f{r_+}{r}\rt)\lt(1-\f{Q^2}{r_+r}\rt). \n
\end{align}
To avoid naked singularity, $r_+$ must be positive. This condition leads to a constraint $M>Q$, so $r_+>Q$, with $Q$ being the event horizon radius of the extremal black hole. Moreover, the Bekenstein--Hawking entropy is one quarter of the event horizon area,
\begin{align}
S=\pi r_+^2, \n
\end{align}
and the Hawking temperature can be calculated as
\begin{align}
T_{\rm H}=\f{f'(r_+)}{4\pi}=\f{1}{4\pi r_+}\lt(1-\f{Q^2}{r_+^2}\rt). \n
\end{align}

Now, we consider black hole thermodynamics in a cavity in asymptotically flat space. In the following discussion on the HP phase transition, we will work in a grand canonical ensemble, in which the cavity plays a role of reservoir, with fixed temperature $T$ and electric potential $\Phi$ on its wall located at radius $r_B$ \cite{Braden}. Naturally, $r_B$ should be larger than the event horizon radius $r_+$, so we have the following constraint,
\begin{align}
Q<r_+<r_B. \n
\end{align}
When the charged black hole is enclosed in such a cavity, its Hawking temperature $T_{\rm H}$ is related to $T$ as
\begin{align}
T =\f{T_{\rm H}}{\sqrt{f(r_B)}}=\f{1-\f{Q^2}{r_+^2}}{4\pi r_+\sqrt{\lt(1-\f{r_+}{r_B}\rt)\lt(1-\f{Q^2}{r_+r_B}\rt)}}, \label{T}
\end{align}
meaning that the temperature $T$ measured at $r_B$ is blue-shifted from the temperature $T_{\rm H}$ measured at infinity by a factor $1/\sqrt{f(r_B)}$. Also, the electric potential $\Phi$ is
\begin{align}
\Phi=\f{A_t(r_B)-A_t(r_+)}{\sqrt{f(r_B)}}=\f{Q\lt(1-\f{r_+}{r_B}\rt)}{r_+\sqrt{\lt(1-\f{r_+}{r_B}\rt)\lt(1-\f{Q^2}{r_+r_B}\rt)}}. \label{Phi}
\end{align}
Furthermore, the thermal energy $E$ of the charged black hole in a cavity can be obtained via the Euclidean action method as \cite{Braden}
\begin{align}
E &= r_B\lt[1-\sqrt{f (r_B)}\rt]= r_B\lt[1-\sqrt{\lt(1-\f{r_+}{r_B}\rt)\lt(1-\f{Q^2}{r_+r_B}\rt)}\rt]. \label{E}
\end{align}

In previous works, the cavity radius $r_B$ was always considered as a constant, so the cavity acts merely as a fixed background, and the $p$--$V$ term is still absent in black hole thermodynamics. Recently, in Ref. \cite{Wang}, the authors introduced an effective volume $V$ for the black hole in a cavity,
\begin{align}
V=\f{4\pi}{3}r^3_B, \label{V}
\end{align}
and allowed $r_B$ to vary. Thus, $V$ becomes a thermodynamic variable, and an effective pressure $p$ can be further defined as its conjugate quantity,
\begin{align}
p=-\lt(\f{\p E}{\p V}\rt)_{S}=\f{1}{4\pi r_B^2} \lt[\f{2-\f{r_+}{r_B}-\f{Q^2}{r_+r_B}}{2\sqrt{\lt(1-\f{r_+}{r_B}\rt)\lt(1-\f{Q^2}{r_+r_B}\rt)}}-1\rt]. \label{p}
\end{align}
In this way, the missing $p$--$V$ term is restored in a new extended phase space, complementary to that in the AdS space. In this new extended phase space, a direct differentiation of Eq. (\ref{E}) yields the first law of black hole thermodynamics \cite{Wang},
\begin{align}
\dd E = T\,\dd S - p\,\dd V+\Phi\,\dd Q. \n
\end{align}
Moreover, a scaling argument directly leads to the corresponding Smarr relation as $E=2TS-3pV+\Phi Q$ \cite{Wang}. Therefore, the analogy between traditional thermodynamics and black hole thermodynamics in the new extended phase space in a cavity is as complete as that in the AdS space.

Equation (\ref{p}) can also be viewed as the equation of state of the black hole in a cavity. In principle, from Eqs. (\ref{T}) and (\ref{V}), solving $r_+$ and $r_B$ in terms of $T$ and $V$, we may arrive at its usual form $p=p(V,T)$. Unfortunately, Eq. (\ref{T}) is an algebraic equation about $r_+$ of higher degree and is analytically unsolvable, so it is impossible to show the explicit form of $p(V,T)$. However, we may investigate two limiting cases: the Schwarzschild black hole with $Q\to 0$ and the extremal black hole with $Q\to M$. Under these circumstances, the analytical forms of $p(V,T)$ do exist, and these results can be used for further reference.

First, for the Schwarzschild black hole in a cavity with $Q\to 0$, we obtain two equations of state (i.e., the large or small black hole branch with larger or smaller $r_+$, respectively),
\begin{align}
p&=\f{1}{4\pi}\lt(\f{4\pi}{3V}\rt)^{2/3}\lt[\f{5-2A}{2\sqrt{6(1-A)}}-1\rt], \label{lbheos} \\
p&=\f{1}{4\pi}\lt(\f{4\pi}{3V}\rt)^{2/3}\lt[\f{5+A-\sqrt{3(1-A^2)}}{2\sqrt{6+3A-3\sqrt{3(1-A^2)}}}-1\rt]. \label{sbheos}
\end{align}
where $4A^3-3A=1-27a/2$ and $a=1/(4\pi T\sqrt[3]{3V/4\pi})^2=(r_+/r_B)^2(1-r_+/r_B)$. In the limit $r_+/r_B\to 1$, we have $a\to 0$ and $A\to 1-3a/2$, and Eq. (\ref{lbheos}) reduces to
\begin{align}
p=\f{T}{2(3V/4\pi)^{1/3}}=\f{T}{v}, \n
\end{align}
where $v=2r_B=2(3V/4\pi)^{1/3}$ is the specific volume of the black hole. This result indicates that the large Schwarzschild black hole in a cavity behaves quite like an ideal gas in the extended phase space, exactly the same as the case in the AdS space. Similarly, in the limit $r_+/r_B\to 0$, Eq. (\ref{sbheos}) reduces to
\begin{align}
p=\f{1}{512\pi^3T^2(3V/4\pi)^{4/3}}=\f{1}{32\pi^3T^2v^4}. \n
\end{align}
It is easy to find that the small black hole has negative heat capacity, so it cannot establish thermal equilibrium with the environment and is thus unstable.

Second, for the extremal black hole in a cavity with $Q\to M$ and $r_+\to Q$, we obtain its equation of state as
\begin{align}
p=\f{\pi Q^4T^2}{2(3V/4\pi)^{4/3}}=\f{8\pi Q^4T^2}{v^4}. \n
\end{align}
This result also resembles that of a non-ideal fluid.

Last, the critical point of the charged black hole in a cavity is determined by
\begin{align}
\lt(\f{\p p}{\p v}\rt)_T=\lt(\f{\p^2 p}{\p v^2}\rt)_T=0. \n
\end{align}
From Eqs. (\ref{T}), (\ref{V}), and (\ref{p}), the critical pressure, special volume, and temperature can be obtained numerically as
\begin{align}
p_{\rm c}=\f{0.0115}{Q^2}, \quad v_{\rm c}=4.14Q, \quad T_{\rm c}=\f{0.117}{Q}, \n
\end{align}
so the critical ratio of $p_{\rm c} v_{\rm c}/T_{\rm c}$ is 0.405 \cite{Wang}. These values only slightly deviate from those in the AdS space: $p_{\rm c}=1/(96\pi Q^2)$, $v_{\rm c}=2\sqrt{6}Q$, $T_{\rm c}=1/(3\sqrt{6}\pi Q)$, and $p_{\rm c} v_{\rm c}/T_{\rm c}=3/8$ \cite{KM}.

\subsection{HP phase transition}

Due to the Hawking mechanism \cite{rad}, a black hole may not only absorb but also radiate energy to the environment. This process will establish thermal equilibrium between the stable large black hole and the surrounding thermal gas, and the phase transition in this black hole--thermal gas system is the HP phase transition. Before detailed discussion, an important point should be made clear. Because of the conservation of charge, a black hole with fixed charge cannot undergo the HP phase transition to the neutral thermal gas. As a result, the HP phase transition should be considered in a grand canonical ensemble, in which the electric potential $\Phi$ is fixed on the wall of cavity and the electric charge $Q$ is allowed to vary, so the thermodynamic potential of interest is the Gibbs free energy $G$.

On the one hand, the Gibbs free energy of the charged black hole is constructed as
\begin{align}
G=E+pV-TS-\Phi Q. \n
\end{align}
Substituting all the relevant expressions in Sect. \ref{sec:excavity} into $G$, we obtain
\begin{align}
G=r_B\lt[\f{7\f{r_+}{r_B}+\f{Q^2}{r_+r_B}-8}{12\sqrt{\lt(1-\f{r_+}{r_B}\rt)\lt(1-\f{Q^2}{r_+r_B}\rt)}}+\f23\rt]. \label{G}
\end{align}
(Here, $G$ is still expressed in terms of $Q$, and its explicit form as a function of $\Phi$ will be shown later in Sect. \ref{sec:RN}.) On the other hand, because the total number of particles of the thermal gas is not conserved but varies with temperature, its Gibbs free energy is always vanishing. In Sect. \ref{sec:HP}, it will be shown that the Gibbs free energy of the black hole monotonically decreases with temperature. Therefore, at low temperatures, the thermal gas phase with vanishing $G$ is more stable; at high temperatures, the black hole phase with negative $G$ is globally preferred, and the thermal gas will collapse into it. Consequently, the criterion of the HP phase transition in the black hole--thermal gas system is
\begin{align}
G =0, \label{cri}
\end{align}
and the HP temperature $T_{\rm HP}$ can be fixed accordingly.

\section{HP phase transitions of the black holes in a cavity} \label{sec:HP}

Below, we study the HP transitions of the Schwarzschild and charged black holes in the extended phase space in a cavity. We focus on three aspects: the HP temperature $T_{\rm HP}$, the minimum black hole temperature $T_0$, and the Gibbs free energy $G$. All analytical results will be shown if possible, and the similarities and dissimilarities between the cavity and AdS cases will also be discussed when needed. For simplicity, we only constrain our attention to the situation $T>T_{\rm c}$; that is to say, we only consider the HP phase transitions, not the van der Waals-like phase transitions below the critical temperature $T_{\rm c}$.

\subsection{HP phase transition of the Schwarzschild black hole in a cavity} \label{sec:Sch}

From Eqs. (\ref{T}), (\ref{p}), and (\ref{G}), the temperature, pressure, and Gibbs free energy of the Schwarzschild black hole in a cavity reduce to
\begin{align}
T &=\f{1}{4\pi r_+ \sqrt{1-\f{r_+}{r_B}}}, \label{TSch}\\
p&=\f{1}{4\pi r_B^2}\lt(\f{2-\f{r_+}{r_B}}{2\sqrt{1-\f{r_+}{r_B}}}-1\rt), \label{pSch}\\
G& =r_B\lt(\f{7\f{r_+}{r_B}-8}{12\sqrt{1-\f{r_+}{r_B}}}+\f23\rt). \label{GSch}
\end{align}
In principle, from Eqs. (\ref{TSch})--(\ref{GSch}), it is straightforward to obtain the explicit form of $G(T,p)$, but the full expression is unnecessarily tedious, and there is no need to show it. Therefore, we will face a series of parametric equations in terms of $r_+$ and $r_B$.

\begin{figure}[h]
\centering
\includegraphics[width=0.6\textwidth]{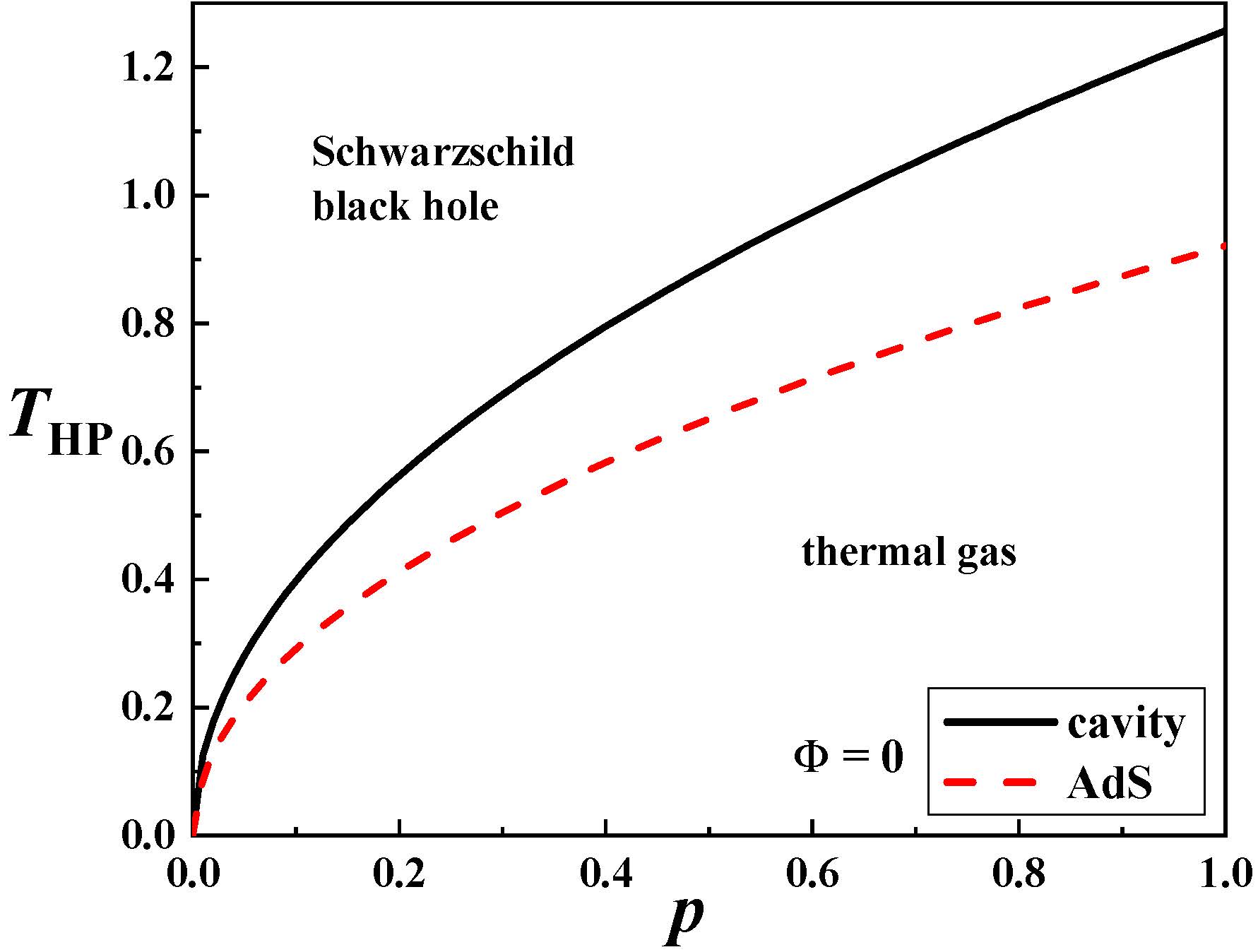}
\caption{\label{fig:TpSch} The coexistence lines for the Schwarzschild black hole in a cavity and in the AdS space. There is no terminal point in the $T_{\rm HP}$--$p$ curve, and the HP phase transition can occur at all pressures. The thermal gas phase lies below the coexistence line, behaving like a solid. The HP temperature $T_{\rm HP}$ increases with $p$ and is a little higher in a cavity than in the AdS space.}
\end{figure}

First, at the HP phase transition point, solving the criterion in Eq. (\ref{cri}) yields
\begin{align}
r_B =\f{49}{48}r_+. \label{4948}
\end{align}
Substituting Eq. (\ref{4948}) into Eqs. (\ref{TSch}) and (\ref{pSch}), we have
\begin{align}
T =\f{7}{4\pi r_+}, \quad p =\f{10368}{16807\pi r_+^2}, \n
\end{align}
so the HP temperature $T_{\rm HP}$ as a function of pressure $p$ (i.e., the equation of coexistence line) is obtained as
\begin{align}
T_{\rm HP} =\f{343}{288}\sqrt{\f{7p}{2\pi}}=1.257\sqrt{p}. \label{TpSch}
\end{align}
This result was also shown numerically in Ref. \cite{Wang}. Moreover, it is analogous to that in the AdS space \cite{HPwo},
\begin{align}
T_{\rm HP}=\sqrt{\f{8p}{3\pi}}=0.921\sqrt{p}. \n
\end{align}
These results confirm the similarity between black hole thermodynamics in a cavity and in the AdS space. Both coexistence lines in these two extended phase spaces are shown in Fig. \ref{fig:TpSch}. From Eq. (\ref{TpSch}), there is no terminal point in the $T_{\rm HP}$--$p$ curve, so the HP phase transition can occur at all pressures, with no critical point. Hence, it is more like a solid--liquid phase transition, rather than a liquid--gas phase transition, and the thermal gas phase even plays a role of solid \cite{1404.2126}, as it is thermodynamically preferred at low temperatures and lies below the coexistence line in Fig. \ref{fig:TpSch}.

Next, we study the phase structure of the Schwarzschild black hole in the HP phase transition. Here, we should first point out a minimum temperature $T_0$, above which there are large and small black hole solutions but below which there is none. To calculate $T_0$, we solve $r_B$ in terms of $r_+$ from Eq. (\ref{TSch}),
\begin{align}
r_B =\f{u^2}{(u+1)(u-1)}r_+, \n
\end{align}
where $u=4\pi r_+T$. Substituting $r_B$ into Eq. (\ref{pSch}), we have
\begin{align}
\f{p}{2\pi T^2}=\f{(u+1)^2(u-1)^4}{u^7}=f(u). \n
\end{align}
It is direct to find the maximum point of $f(u)$ is at $u =1+2\sqrt{2}$, and $f(u)\leq\f{256(1+\sqrt{2})^2}{(1+2\sqrt{2})^7}$. Therefore, the black hole temperature has a lower bound $T_0$ as
\begin{align}
T\geq T_0=\f{(1+2\sqrt{2})^{7/2}}{16(1+\sqrt{2})}\sqrt{\f{p}{2\pi}}=1.134\sqrt{p}. \n
\end{align}
This result is also consistent with the numerical value in Ref. \cite{Wang}. Meanwhile, the minimum temperature of the Schwarzschild black hole in the AdS space is \cite{HPwo}
\begin{align}
T\geq T_0= \sqrt{\f{2p}{\pi}}=0.798\sqrt{p}. \n
\end{align}
We observe that both $T_{\rm HP}$ and $T_0$ of the Schwarzschild black hole in a cavity are higher than the AdS counterparts.

With the above preparations, we are able to investigate the Gibbs free energy of the black hole--thermal gas system. As already mentioned, we utilize the simple parametric equations for $T$ and $G$ instead of the lengthy explicit expression. For this purpose, we first solve $r_+$ from Eq. (\ref{pSch}),
\begin{align}
r_+=xr_B, \label{r+rBSch}
\end{align}
where
\begin{align}
x=4\sqrt{2\pi pr_B^2(2\pi pr_B^2+1)}\Big(\sqrt{2\pi pr_B^2+1}-\sqrt{2\pi pr_B^2}\Big)^2. \n
\end{align}
It is easy to find $x$ is less than 1, so $r_+<r_B$ is always ensured for the Schwarzschild black hole in a cavity (Attention, this is not the case for the charged black hole in a cavity, see Sect. \ref{sec:RN}). Substituting Eq. (\ref{r+rBSch}) into Eqs. (\ref{TSch}) and (\ref{GSch}), we can reexpress $T$ and $G$ in terms of $r_B$,
\begin{align}
T &=\f{1}{4\pi r_B x\sqrt{1-x}}=T(p,r_B), \label{GTTSch}\\
G &=r_B\lt(\f{7x-8}{12\sqrt{1-x}}+\f23\rt)=G(p,r_B). \label{GTGSch}
\end{align}

\begin{figure}[h]
\centering
\includegraphics[width=0.6\textwidth]{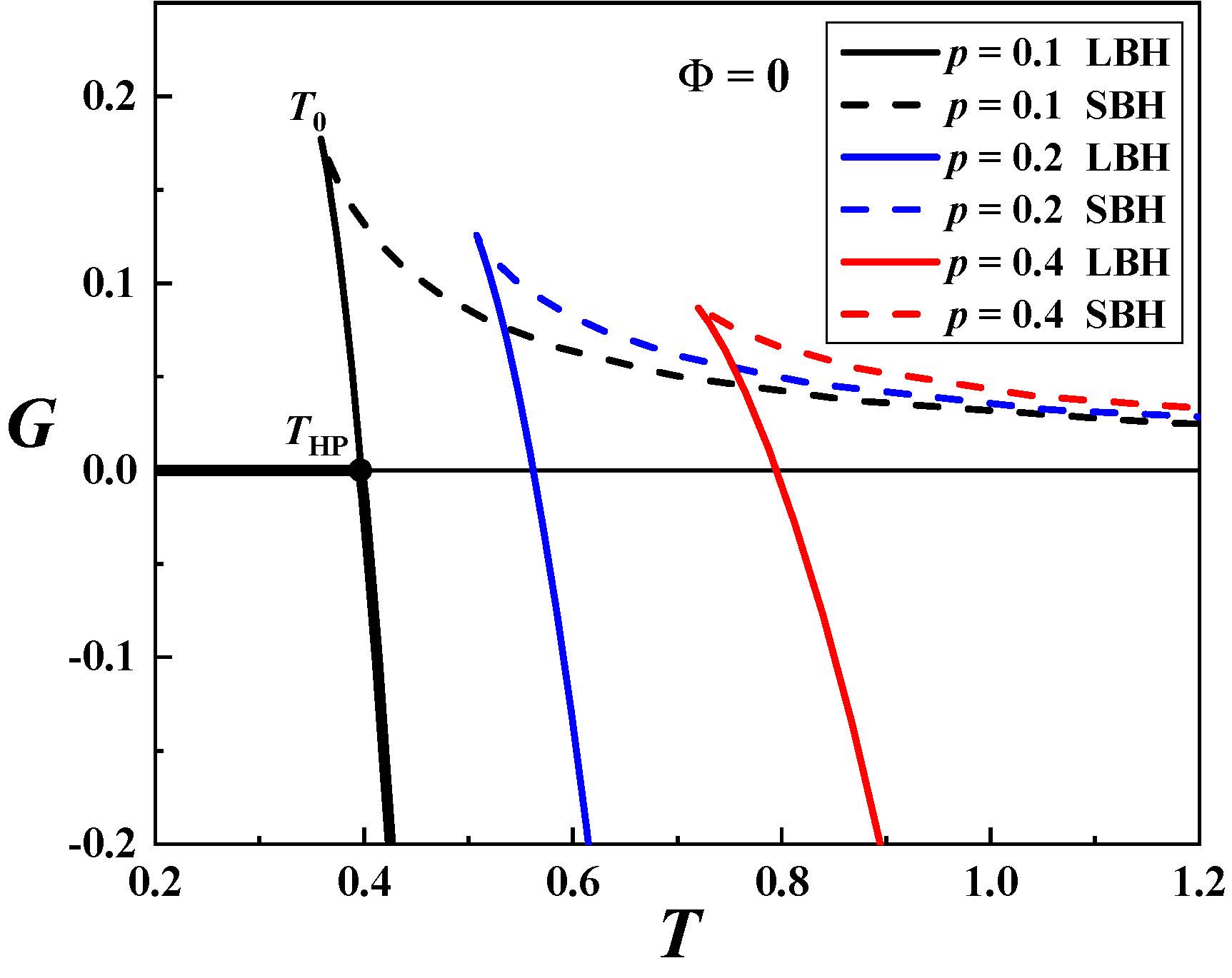}
\caption{\label{fig:GTSch} The Gibbs free energy of the black hole--thermal gas system in a cavity, with different values of pressure $p$. The $G$--$T$ curves of the thermal gas and the stable large black holes intersect at the HP temperature $T_{\rm HP}$, which increases at large pressures. Below or above $T_{\rm HP}$, the thermal gas phase or the large black hole phase is globally preferred. The $G$--$T$ curves of the unstable small black holes are concave and always above the $T$-axis, so the HP phase transition never occurs. (LBH and SBH stand for large and small black hole, respectively.)}
\end{figure}

By means of these parametric equations, we plot the $G$--$T$ curves of the Schwarzschild black holes in a cavity in Fig. \ref{fig:GTSch}, with different pressures [these curves can also be plotted via Eqs. (\ref{lbheos}) and (\ref{sbheos})]. We observe two branches of the $G$--$T$ curves, corresponding to the large and small black holes, and these two curves meet with a cusp at the minimum black hole temperature $T_0$, below which no black hole solution exists any longer. Both Gibbs free energies of the large and small black holes decrease with temperature. However, from the heat capacity at constant pressure, $C_p=T(\p S/\p T)_p=-T(\p^2 G/\p T^2)_p$, the small black holes have negative $C_p$ (their $G$--$T$ curves are always concave), so they are unstable and cannot maintain thermal equilibrium with the environment. Moreover, their $G$--$T$ curves only tend to but never reach the $T$-axis, so the HP phase transition never occurs. On the contrary, the large black holes are stable, and their $G$--$T$ curves are convex and cross the $T$-axis at the HP temperature $T_{\rm HP}$. With $p$ increasing, $T_{\rm HP}$ moves rightward, meaning that $T_{\rm HP}$ increases at high pressures, consistent with the coexistence line in Fig. \ref{fig:TpSch}. Furthermore, below $T_{\rm HP}$, the thermal gas phase with vanishing Gibbs free energy is more stable, and above $T_{\rm HP}$, the large black hole phase with negative Gibbs free energy is thermodynamically preferred. Altogether, there is a discontinuity in the slopes of the $G$--$T$ curves of the black hole--thermal gas system, corresponding to the first-order HP phase transition at $T_{\rm HP}$.

\begin{figure}[h]
\centering
\includegraphics[width=0.6\textwidth]{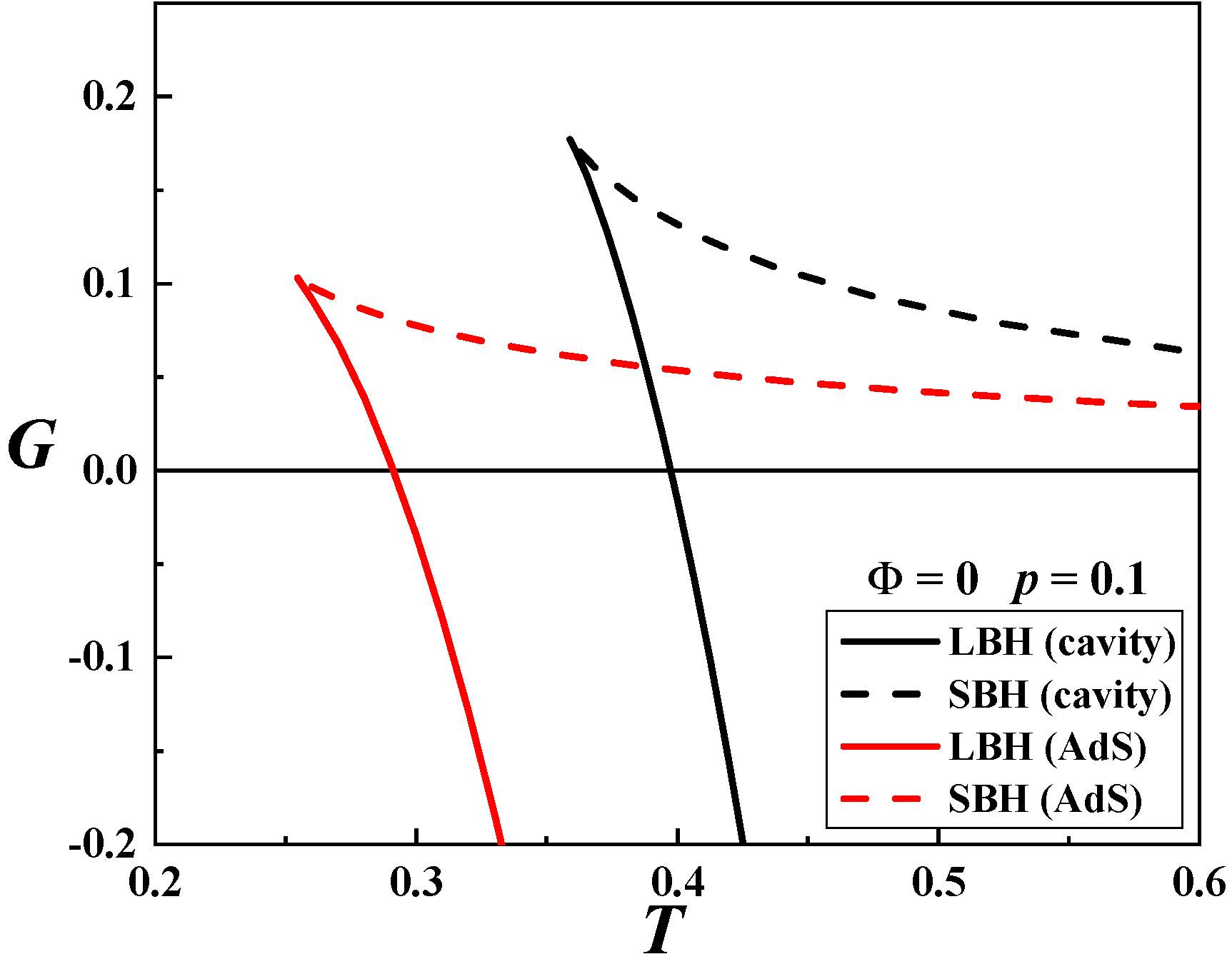}
\caption{\label{fig:GTcavAdS} The Gibbs free energy of the Schwarzschild black holes in a cavity and in the AdS space. Both the HP temperature $T_{\rm HP}$ and the minimum black hole temperature $T_0$ in the cavity case are higher than those in the AdS case at the same pressure.}
\end{figure}

For comparison, we also plot the $G$--$T$ curves of the Schwarzschild black holes in a cavity and in the AdS space together in Fig \ref{fig:GTcavAdS}. The shapes of these curves are similar, only with the cavity case having higher $T_{\rm HP}$ and $T_0$ than the AdS case at the same pressure. This observation further verifies the analogy between these two extended phase spaces.

\subsection{HP phase transition of the charged black hole in a cavity} \label{sec:RN}

Now, we continue to discuss the HP phase transition of the charged black hole in a cavity and compare the relevant results with those in the AdS counterparts. Generally speaking, the following procedure is in parallel to that in Sect. \ref{sec:Sch}, but as it will be shown, there are several intrinsic differences in between. First, we will work in the grand canonical ensemble with fixed electric potential $\Phi$. Second, if the HP phase transition occurs, $\Phi$ must have an upper bound, and the charged black hole with $\Phi$ larger than this bound cannot have the HP phase transition. Third, due to the constraint $r_+<r_B$, the large black hole temperature also has an upper bound (i.e., its $G$--$T$ curve has a terminal point). All these issues will be explained in depth below.

We start our discussion by reexpressing everything in terms of the electric potential $\Phi$ instead of the charge $Q$. From Eq. (\ref{Phi}), we have
\begin{align}
Q =\f{\Phi r_+}{\sqrt{1-(1-\Phi^2)\f{r_+}{r_B}}}, \n
\end{align}
and it is not hard to see the upper bound of $\Phi$,
\begin{align}
\Phi<\f{Q}{r_+}<1. \label{Phi1}
\end{align}
Substituting $Q$ into Eqs. (\ref{T}), (\ref{p}), and (\ref{G}), we obtain the temperature, pressure, and Gibbs free energy of the charged black hole in a cavity in terms of $\Phi$,
\begin{align}
T &=\f{1-\Phi^2}{4\pi r_+ \sqrt{1-(1-\Phi^2)\f{r_+}{r_B}}}, \label{TRN}\\
p&=\f{1}{4\pi r_B^2}\lt[\f{2-(1-\Phi^2)\f{r_+}{r_B}}{2\sqrt{1-(1-\Phi^2)\f{r_+}{r_B}}}-1\rt],\label{pRN}\\
G &=r_B\lt[\f{7(1-\Phi^2)\f{r_+}{r_B}-8 }{12\sqrt{1-(1-\Phi^2)\f{r_+}{r_B} }}+\f 23\rt]. \label{GRN}
\end{align}
Naturally, these results reduce to those in Eqs. (\ref{TSch})--(\ref{GSch}) when $\Phi$ vanishes.

Again, at the HP phase transition point, from the criterion in Eq. (\ref{cri}), we have
\begin{align}
r_B =(1-\Phi^2)\f{49}{48}r_+. \label{rB}
\end{align}
Taking into account the constraint $r_+<r_B$, we find that, if the HP phase transition occurs, $\Phi$ must have a much tighter upper bound than that in Eq. (\ref{Phi1}),
\begin{align}
\Phi<\f 17. \n
\end{align}
In other words, when $1/7<\Phi<1$, even if there is the charged black hole solution in a cavity, the HP phase transition cannot occur. This is quite different from the case of the Schwarzschild black hole in Sect. \ref{sec:Sch}, where the HP phase transition always occurs.

\begin{figure}[h]
\centering
\includegraphics[width=0.6\textwidth]{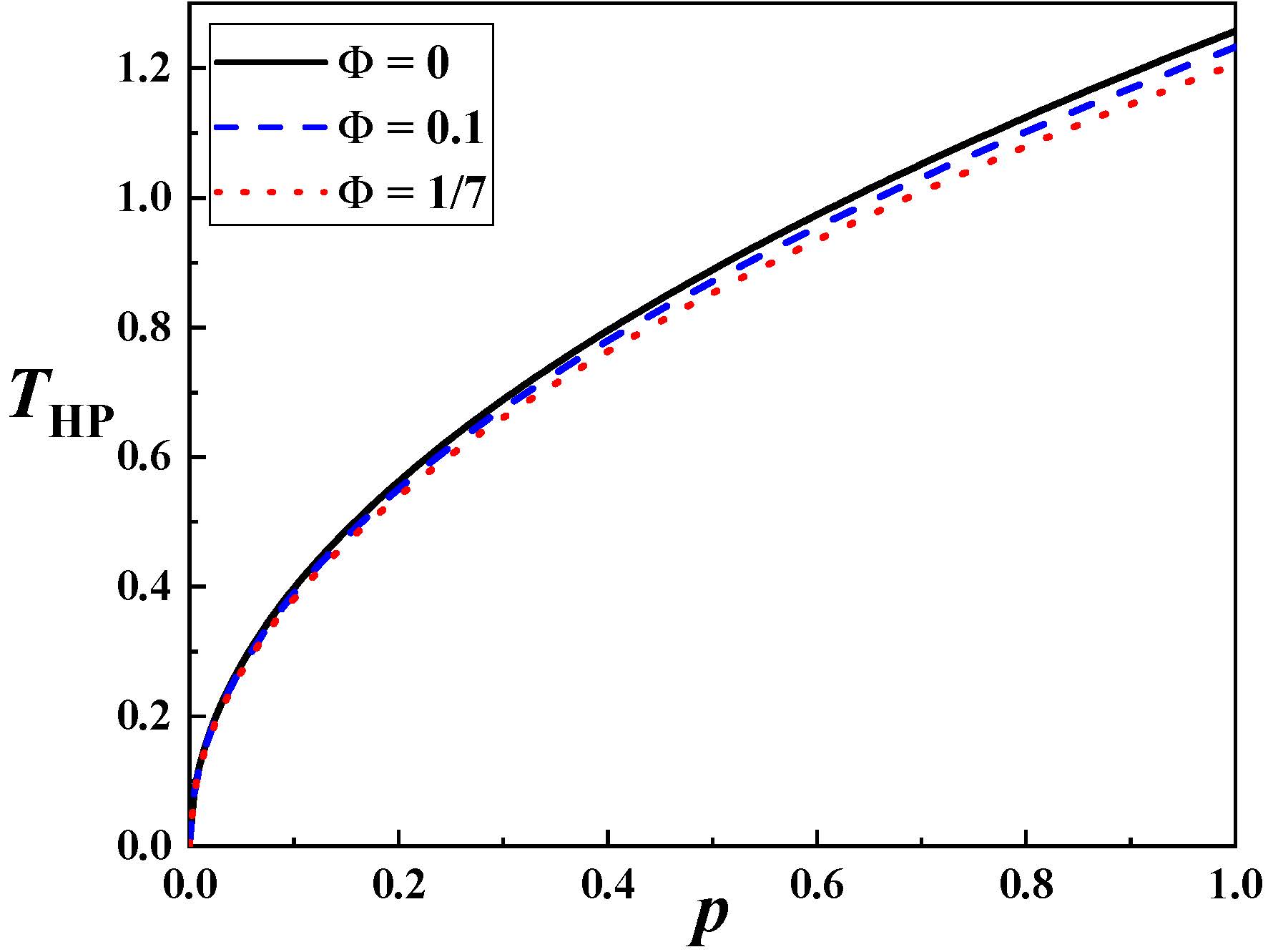}
\caption{\label{fig:TpRN} The coexistence lines for the charged black hole in a cavity, with different values of electric potential $\Phi$. The HP phase transition can occur at all pressures, and $T_{\rm HP}$ decreases with $\Phi$ at a fixed pressure. $\Phi$ has an upper bound of $1/7$, above which there is no HP phase transition.}
\end{figure}

Substituting Eq. (\ref{rB}) into Eqs. (\ref{TRN}) and (\ref{pRN}), we obtain the equation of coexistence line for the charged black hole in a cavity,
\begin{align}
T_{\rm HP} =(1-\Phi^2)^2\f{343}{288}\sqrt{\f{7p}{2\pi}}. \label{TpRN}
\end{align}
Compared to the result in the AdS space \cite{HPwo},
\begin{align}
T_{\rm HP}=\sqrt{\f{(1-\Phi^2)8p}{3\pi}}, \n
\end{align}
the HP temperature $T_{\rm HP}$ decreases with $\Phi$ even faster. The coexistence lines for the charged black hole in a cavity are shown in Fig. \ref{fig:TpRN}, with different electric potentials.

Next, the minimum black hole temperature $T_0$ can also be obtained by the same method in Sect. \ref{sec:Sch},
\begin{align}
T\geq T_0=(1-\Phi^2)^2\f{(1+2\sqrt{2})^{7/2}}{16(1+\sqrt{2})}\sqrt{\f{p}{2\pi}}, \label{T0RN}
\end{align}
and $T_0$ is modified with the same factor $(1-\Phi^2)^2$ as that in $T_{\rm HP}$ in Eq. (\ref{TpRN}).

Last, we explore the Gibbs free energy for the charged black hole--thermal gas system. Again, we first solve $r_+$ from Eq. (\ref{pRN}),
\begin{align}
r_+=\f{x}{1-\Phi^2}r_B. \label{r+rBRN}
\end{align}
However, before substituting $r_+$ into $T$ and $G$, we should emphasize an important point in advance. Different from Eq. (\ref{r+rBSch}), where $r_+=xr_B$ is always less than $r_B$, now due to the factor $1-\Phi^2$ in the denominator, we must set a new constraint as
\begin{align}
x&=4\sqrt{2\pi pr_B^2(2\pi pr_B^2+1)}\Big(\sqrt{2\pi pr_B^2+1}-\sqrt{2\pi pr_B^2}\Big)^2<1-\Phi^2, \label{ys1}
\end{align}
such that $r_+<r_B$. Since $x$ is a monotonically increasing function of $r_B$, solving the above inequality, we find that, in the grand canonical ensemble with fixed electric potential, the cavity radius $r_B$ has an upper bound as
\begin{align}
r_B<\f{1-\Phi}{2\sqrt{2\pi p \Phi}}. \label{ys2}
\end{align}
This result is significantly different from the Schwarzschild case, where $r_B$ can take any value. Of course, in the limit $\Phi\to 0$, the upper bound of $r_B$ goes to infinity.

\begin{figure}[h]
\centering 
\includegraphics[width=.45\textwidth]{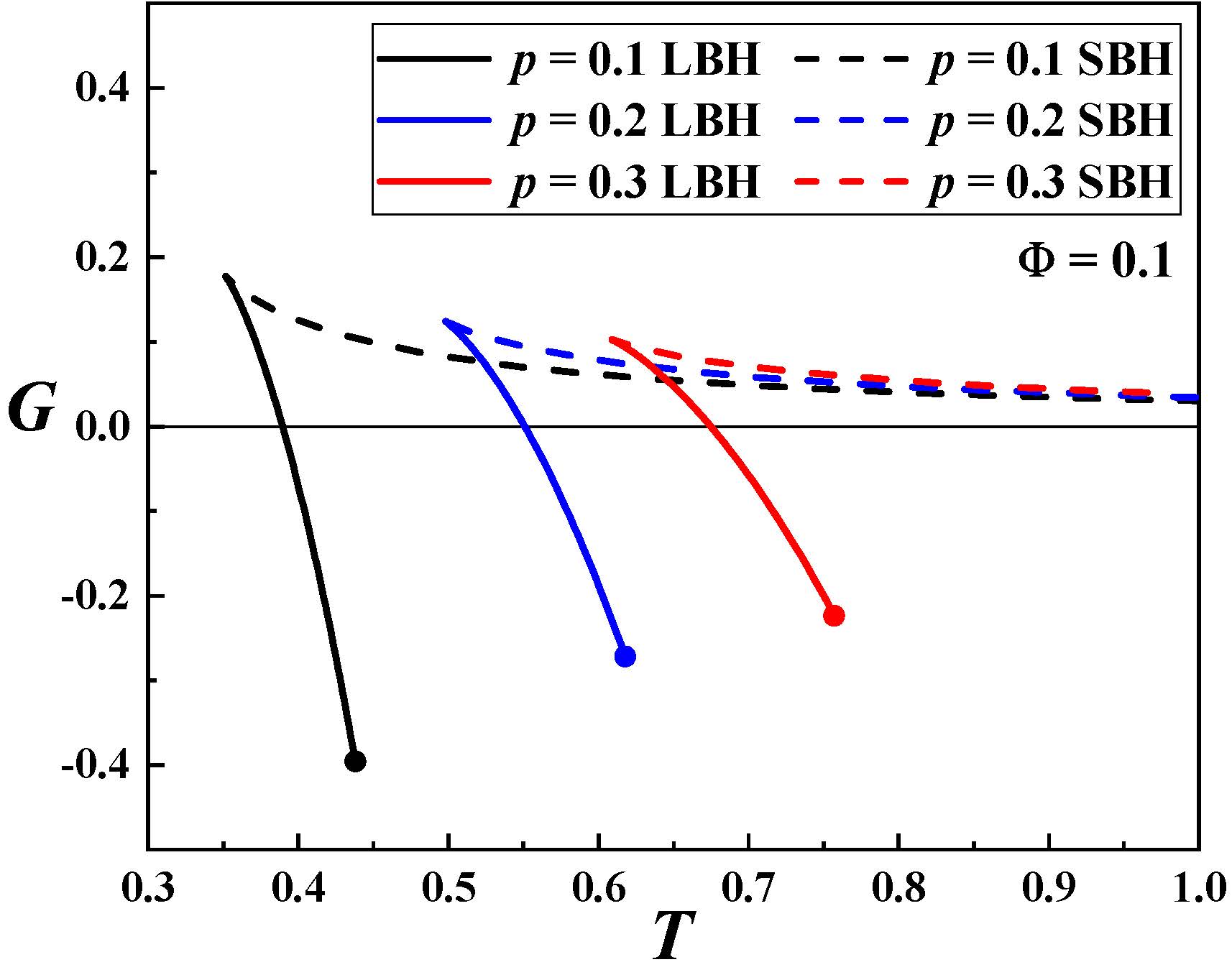} \quad 
\includegraphics[width=.46\textwidth]{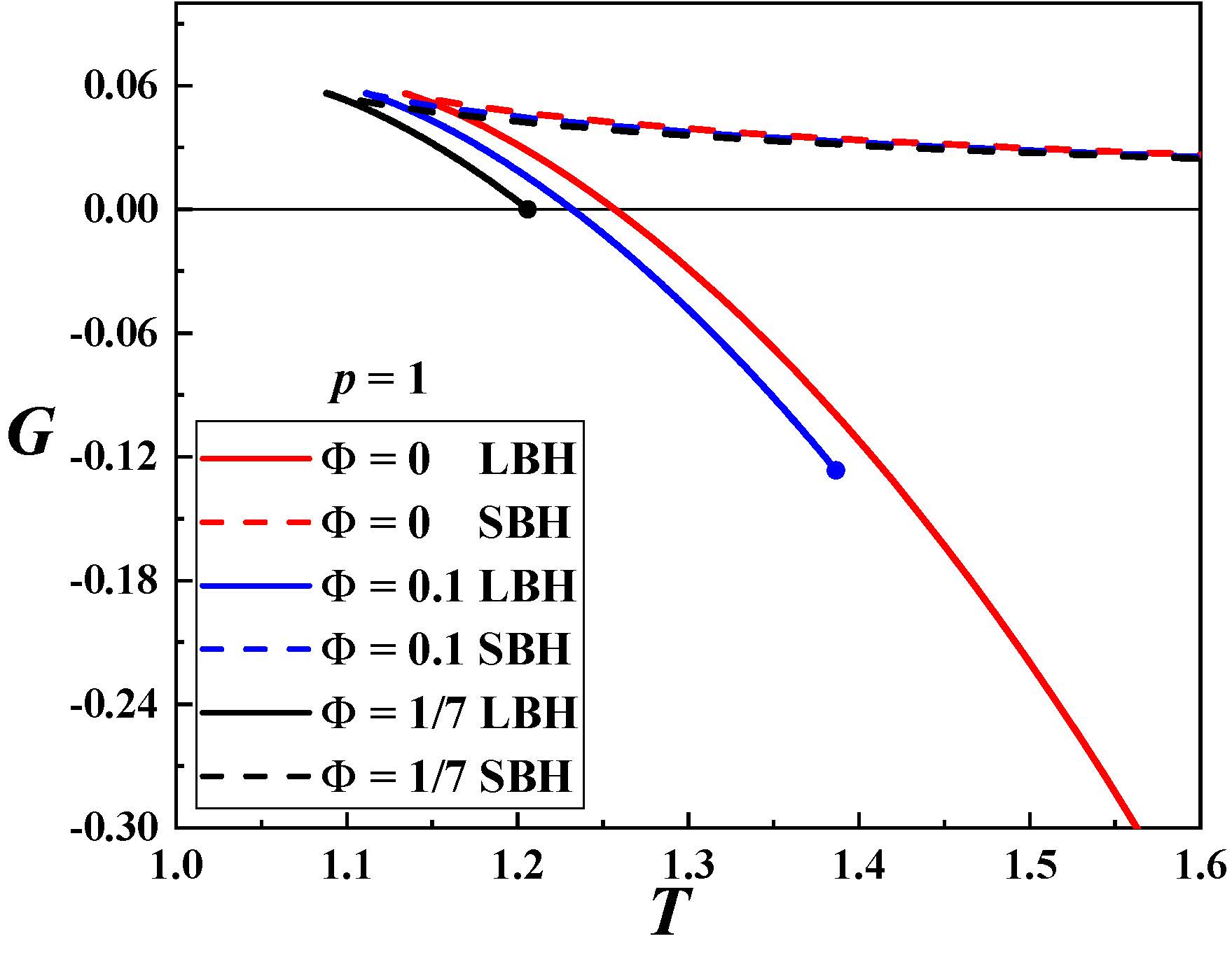}
\caption{\label{fig:GTRN} The Gibbs free energy of the charged black hole--thermal gas system in a cavity, with different values of pressure $p$ and electric potential $\Phi$. The HP temperature $T_{\rm HP}$ increases with $p$ and decreases with $\Phi$. The small charged black holes have no terminal point in their $G$--$T$ curves and also no HP phase transition. The large charged black holes have the HP phase transitions, but also have the terminal points in their $G$--$T$ curves. In the right panel, when $\Phi=0$, the $G$--$T$ curve extends to infinity; when $\Phi$ increases, the terminal points move up left; when $\Phi\to 1/7$, the terminal point is located just on the $T$-axis at $(T_{\rm HP},0)$.}
\end{figure}

Substituting Eq. (\ref{r+rBRN}) into Eqs. (\ref{TRN}) and (\ref{GRN}), we can reexpress $T$ and $G$ in terms of $r_B$,
\begin{align}
T &=\f{(1-\Phi^2)^2}{4\pi r_B x\sqrt{1-x}}=T(p,r_B), \label{GTTRN} \\
G &=r_B\lt(\f{7x-8}{12\sqrt{1-x}}+\f23\rt)=G(p,r_B). \label{GTGRN}
\end{align}
From Eq. (\ref{GTTRN}), we find that $T$ is modified by the same factor $(1-\Phi^2)^2$ as $T_{\rm HP}$ and $T_0$. Meanwhile, from Eq. (\ref{GTGRN}), $G$ remains the same expression as that in the Schwarzschild case in Eq. (\ref{GTGSch}). With these parametric equations, we plot the $G$--$T$ curves of the charged black holes in a cavity in Fig. \ref{fig:GTRN}, with different pressures and electric potentials. We observe that $T_{\rm HP}$ increases with $p$ and decreases with $\Phi$, consistent with Fig. \ref{fig:TpRN}.

Here, we point out a characteristic feature in Fig. \ref{fig:GTRN}. Because of the constraints in Eqs. (\ref{ys1}) and (\ref{ys2}), the $G$--$T$ curves of the large charged black holes in a cavity have the terminal points at
\begin{align}
(T,G)=\lt((1+\Phi)\sqrt{\f{p}{2\pi\Phi}},-\f{(1-\Phi)^2(1-7\Phi)}{24\sqrt{2\pi p\Phi^3}}\rt). \n
\end{align}
This means that the temperature of the large charged black hole has an upper bound. In particular, in the right panel of Fig. \ref{fig:GTRN}, as $\Phi$ increases, the terminal points move toward the upper left corner of the $G$--$T$ plane, and finally when $\Phi\to 1/7$, the terminal point is located exactly on the $T$-axis at the point $(T_{\rm HP},0)= (\sqrt{32p/7\pi},0)$, which is just the result extracted from Eq. (\ref{TpRN}). Also, when $\Phi\to 0$, there is no terminal point in the $G$--$T$ curve, as already shown for the Schwarzschild black hole in a cavity in Fig. \ref{fig:GTSch}.

We need not repeat the comparison with the AdS case, and the situation is rather similar as that in Fig. \ref{fig:GTcavAdS}. The unique difference is that there is no terminal point in the $G$--$T$ curves of the large charged black holes in the AdS case, as the event horizon radius $r_+$ can take any value, without constraint like $r_+<r_B$ in a cavity.

\section{Conclusion} \label{sec:con}

From the thermodynamic perspective, a black hole resembles a complicated system with temperature and entropy. However, a black hole in asymptotically flat space has negative heat capacity and is thermodynamically unstable. This instability can be avoided by imposing appropriate boundary conditions. By doing so, the effective pressure and volume can also be introduced simultaneously, and such theories are thus named as black hole thermodynamics in the extended phase space. There can be two complementary approaches to extend the black hole phase space. One is to enclose the black hole in the AdS space and to introduce the effective pressure via the varying cosmological constant $\Lambda$ \cite{KM}, and the other is to confine the black hole in a cavity in asymptotically flat space and to introduce the effective volume via the varying cavity radius $r_B$ \cite{Wang}. The former has received great interest in recent years, but the relevant studies on the latter are just starting.

In our present work, we investigate one of the most important issues in black hole thermodynamics: the HP phase transitions between the thermal gas and the black holes in a cavity. Our basic conclusions can be drawn as: (1) there can be large and small black hole solutions in a cavity, when the black hole temperature is above a minimum temperature $T_0$. (2) The Gibbs free energies of the small black holes are always positive, so there is no HP phase transition totally. The Gibbs free energies of the large black holes decrease with temperature, and there are HP phase transitions at the HP temperature $T_{\rm HP}$. (3) For the Schwarzschild black hole in a cavity, the HP phase transition occurs at all pressures, and $T_{\rm HP}$ increases with pressure $p$. (4) For the charged black hole in a cavity, in the grand canonical ensemble with fixed electric potential $\Phi$, $T_{\rm HP}$ increases with $p$ and decreases with $\Phi$. (5) If the HP phase transition occurs, $\Phi$ has an upper bound of $1/7$, and above this value, no HP phase transition exists, although there are still black hole solutions.

Furthermore, we compare these results to our previous work in the AdS space in Ref. \cite{HPwo} and find remarkable similarities in $T_{\rm HP}$--$p$ and $G$--$T$ curves, both for the Schwarzschild and charged black holes. At the same time, there is a notable difference in between. The $G$--$T$ curves for the charged black holes in the grand canonical ensemble in the AdS space can extend to infinity, but have terminal points in the cavity case, and when $\Phi\to 1/7$, the terminal point lies exactly on the $T$-axis at $(T_{\rm HP},0)$. This particular discrepancy stems from the fact that, when a black hole is in a cavity, its event horizon radius $r_+$ must be smaller than the cavity radius $r_B$. However, in the AdS space, there is no such constraint. Altogether, we intend to provide a whole picture of the HP phase transitions in the two extended phase spaces.

The above observations naturally motivate further investigations on these two extended phase spaces. Of course, they are equivalent in the limits $\Lambda\to 0$ and $r_B\to\infty$. However, at finite values of $\Lambda$ and $r_B$, there still seem to be some issues sensitive to the boundary conditions. Therefore, it would be highly necessary and interesting to explore beyond the present level. For instance, going to higher dimensions is a next step naturally, in which some typical phenomena are expected (e.g., the reentrant phase transitions). Besides, the inclusion of angular momentum of the black hole is also a straightforward generalization. However, from our experience in Ref. \cite{HPwo}, it is not very hopeful to gain much new physical insight due to rotation, except some mathematical complexity. In fact, all the relevant results obtained in the AdS space in the previous literature can be reevaluated in the new extended phase space in a cavity. In this way, we wish to achieve a deeper understanding of black hole thermodynamics in different extended phase spaces with various boundary conditions.

\acknowledgments

We are very grateful to Bing-Yu Su and Peng Wang for fruitful discussions. This work is supported by the Fundamental Research Funds for the Central Universities of China (No. N182410001).


\end{document}